\documentclass[%
 reprint,
superscriptaddress,
 amsmath,amssymb,
 aps,
prl,
floatfix,
]{revtex4-1}

\usepackage{graphicx}
\graphicspath{{figures/}} 
\usepackage{dcolumn}
\usepackage{bm}
\usepackage{physics}
\usepackage{tikz}
\usepackage{xcolor}
\usepackage{siunitx}
\DeclareSIUnit{\molar}{M}
\newcommand{\um}[1]{\SI{#1}{\micro\molar}}
\usepackage{tabularx}
\usepackage{xspace}
\usepackage{comment}
\usepackage[ruled,vlined]{algorithm2e}
\usepackage{algpseudocode}
\newcommand*\tageq{\refstepcounter{equation}\tag{\theequation}}
\newcommand{\lnorm}[1]{\ensuremath{\lVert #1 \rVert}}
\newcommand{\eg}{\textit{e.g.}\xspace}
\usepackage{etoolbox}

\DeclareMathOperator*{\argmin}{argmin}

\newcommand{\boldf}[1]{#1}

\usepackage[colorlinks=true,allcolors=blue]{hyperref}

\newcommand \paper {paper}
\newcommand \supp {Appendix }

\newcommand \brandeis {Department of Physics, Brandeis University, Waltham, Massachusetts 02453, USA
}

\newcommand \tufts {Department of Physics and Astronomy, Tufts University, Medford, Massachusetts 02155, USA}

\newcommand \ucsb {Department of Physics, University of California at Santa Barbara, Santa Barbara, California 93106, USA}

\newcommand \mich {
Department of Physics, University of Michigan, Ann Arbor, Michigan 48109 USA}

\newcommand{\q}{\vb{Q}}
\newcommand{\vel}{\vb{u}}

\newcommand{\ubar}[1]{\text{\b{$#1$}}}

\newcommand*{\nStar}{n^*}

\begin{document}


\title{Data-driven discovery of active nematic hydrodynamics}%

\author{Chaitanya Joshi}
\email{chaitanya.joshi@tufts.edu}
\affiliation{\brandeis}
\affiliation{\tufts}%

\author{Sattvic Ray}
\affiliation{\ucsb}

\author{Linnea M. Lemma}
\affiliation{\brandeis}
\affiliation{\ucsb}

\author{Minu Varghese}
\affiliation{\brandeis}
\affiliation{\mich}

\author{Graham Sharp}
\affiliation{\ucsb}%

\author{Zvonimir Dogic}
\email{zdogic@physics.ucsb.edu}
\affiliation{\ucsb }

\author{Aparna Baskaran}
\email{aparna@brandeis.edu}
\affiliation{\brandeis}%

\author{Michael F. Hagan}
\email{hagan@brandeis.edu}
\affiliation{\brandeis}%

\date{\today}

\begin{abstract}
 Two-dimensional active nematics are often modeled using phenomenological continuum theories that describe the dynamics of the nematic director and fluid velocity through partial differential equations (PDEs). While these models provide a statistically accurate description of the experiments, the identification of the relevant terms in the PDEs and their parameters is usually indirect. Here, we adapt a recently developed method to automatically identify optimal continuum models for active nematics directly from the spatio-temporal director and velocity data, via sparse fitting of the coarse-grained fields onto generic low order PDEs. We test the method extensively on computational models, and then apply it to data from experiments on microtubule-based active nematics. Thereby, we identify the optimal models for microtubule-based active nematics, along with the relevant phenomenological parameters. We find that the dynamics of the orientation field are largely governed by its coupling to the underlying flow, with free-energy gradients playing a negligible role. Furthermore, by fitting the flow equation to experimental data, we estimate a key parameter quantifying the `activity' of the nematic.
\end{abstract}

\pacs{Valid PACS appear here}
\maketitle

Active nematics demonstrate how energy-consuming motile constituents can self-organize into diverse non-equilibrium dynamical states~\cite{Marchetti2013,Ramaswamy2010,Toner2005}. They offer a versatile platform to both advance our fundamental understanding of non-equilibrium physics and develop materials with properties that are thermodynamically forbidden in equilibrium. These twin goals require theoretical models that can reveal the mechanism underlying the emergent dynamics, and guide rational design to elicit desired spatio-temporal dynamics. In this \paper, we combine data-driven model discovery with experiments and computational modeling to identify the most parsimonious model for a specific experimental realization of active nematics. Using the discovered model, we identify the relationship between key theoretical parameters, such as the magnitude of activity, and experimental control variables. The described methods can be applied to diverse realizations of active nematics ranging from shaken rods to motile cells~\cite{Narayan2007,Wensink2012,Zhou2014,Duclos2017,Kawaguchi2017,Kumar2018}, as well as other realizations of active matter.

Our target is to describe microtubule-based active nematics. Being reconstituted from tunable and well-characterized components, they afford a unique opportunity to develop continuum theory models and connect these to microscopic dynamics~\cite{Sanchez2012,DeCamp2015,Doostmohammadi2018}. 
Hydrodynamic theories based on the expansion of symmetry-allowed terms have provided insight into dynamics of active nematics in general, and the microtubule-based system specifically. For example, such models have been used to describe defect dynamics \cite{Giomi2013,Giomi2014b,Doostmohammadi2017,Oza2016,Cortese2018,Shendruk2018}, induced flows in the suspending fluid \cite{Thampi2014,Giomi2015,Lemma2019}, and how confinement in planar \cite{Shendruk2017,Norton2018,Gao2017} and curved geometries \cite{Zhang2016,Ellis2017,Alaimo2017} controls defect proliferation and dynamics. These efforts employed a range of hydrodynamic models, but each assumed different symmetry-allowed terms. Thus, the field lacks a standard model and understanding of magnitudes and sources of error in existing models.

Improving  hydrodynamic theories is a formidable task. Accounting for higher-order effects involves an intractable number of additional symmetry-allowed terms, and the phenomenological coefficient associated with each term must be estimated from experimental data. To overcome these limitations, we identify the most parsimonious model that captures the experimentally observed dynamics. We build on a previous framework~\cite{Brunton2016a,Rudy2017} that has recently been applied to  particle-based simulations of active matter~\cite{Maddu2022}  and computational and experimental data of overdamped polar particles~\cite{Supekar2021}. 
We employ extensive PolScope and fluorescence measurements of microtubule alignment and PIV measurements of velocities, to identify equations governing both the orientational dynamics and the activity-driven flows coupled to long-range hydrodynamic interactions in the fluid momentum equation.

\begin{figure}
    \includegraphics{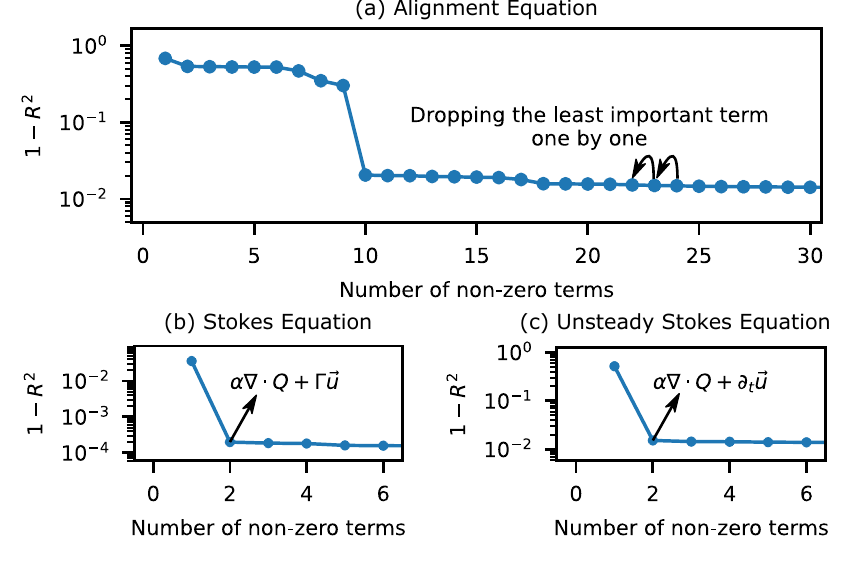}
    \caption{Benchmarking model discovery framework on continuum simulations data. \emph{Optimality curves}: $1-R^2$ as a function of number of non-zero terms in the model for (a) the $Q_{xx}$ component of Eq.~\eqref{eq:Qform}, (b) the Stokes equation and (c) the unsteady Stokes equation. (a): Starting with the full library, the least important term is eliminated one by one to get a hierarchy of models. The $R^2$ deteriorates when terms used in the simulation begin to drop. (b), (c): same as (a) for the flow equation, Eq.~\eqref{eq:uformweak}, excluding the pressure.}
    \label{fig:sindy_sim}
\end{figure}

Our framework is built on the Sparse Identification of Non-linear Dynamics (SINDy) algorithm~\cite{Brunton2016a} and its extension to PDEs (PDE-FIND)~\cite{Rudy2017}. We adapt some key improvements of this method to suit the microtubule-based active nematics system~\cite{Alves2020,Reinbold2019,Reinbold2020,Reinbold2021}. With the orientation and velocity data available from the experiments, we seek models describing the active nematic as a single 2D fluid with nematic symmetry \cite{beris_edwards_1994,Doostmohammadi2018}. Hence, there are two fields:
the nematic tensor order parameter $\vb{Q} = s [\vb{n}\otimes\vb{n}-(1/2)\vb{I}]$ and a flow
field $\vb{u}$, with $\vb{n}$ as the local orientation unit vector and $s$ the scalar order parameter. $\vb{Q}$ is symmetric and trace-less by definition. We assume a constant density and an incompressible fluid; the latter is validated by numerical measurements of the divergence of the velocity field~\cite{Lemma2021}. Our model then consists of 4 independent scalar fields: $Q_{xx}$, $Q_{xy}$, $u_x$, $u_y$, and a latent variable $P$ (pressure). 

We begin by postulating the generalized form of the underlying model. The dynamics of the Q-tensor  takes the form common throughout all continuum theories of active nematics:
\begin{equation}
    \partial_t Q_{ij} = \sum_k a^k_{ij} F_k(\q, \vel,\nabla\q, \nabla\vel, \ldots)
    \label{eq:Qform}
\end{equation}
where the $F_k$'s are combinations (potentially non-linear) of $\vb{Q}$, $\vb{u}$, and their spatial derivatives up to a maximum order, and $a^k_{ij}$'s are the corresponding phenomenological coefficients. 
For instance, in 2D, a well-known model for the Q-equation is \cite{Doostmohammadi2018}:
\begin{equation}
    \partial_t \vb{Q} + \vb{u}\cdot\nabla\vb{Q} - \vb{S} = \Gamma_r \vb{H}
    \label{eq:Qtheory}
\end{equation}
where $\vb{S} = - (\vb{\Omega}\cdot\vb{Q} - \vb{Q}\cdot\vb{\Omega}) + \lambda \vb{E} -2\lambda \vb{Q} (\vb{Q}\colon\nabla\vb{u})$ is the co-rotation term and $\vb{H} = a_2 \vb{Q} + a_4 \Tr(\vb{Q}^2)\vb{Q} + K\nabla^2\vb{Q}$ is the negative gradient of the liquid crystal free energy. Here, $E_{ij} = (\partial_i u_j + \partial_j u_i)/2$ and $\Omega_{ij} = (\partial_i u_j - \partial_j u_i)/2$ are the strain rate and vorticity tensors respectively, $\lambda$ is the flow alignment parameter, $\Gamma_r$ is the rotational diffusion coefficient, $K$ is the bending modulus, and $a_2>0$, $a_4<0$ are phenomenological coefficients corresponding to the isotropic-nematic transition (see \supp Section A for an extended discussion of the theory). We build a vast library of the terms $F_k$ that can capture models well beyond Eq.~\eqref{eq:Qtheory}.
Further, we make no physics-based simplifying assumptions, e.g. translational, rotational, and Galilean invariance, for the alignment equation (Eq.~\ref{eq:Qform}). Hence, the discovery of a model which satisfies these conditions is a test of the algorithm. This results in a large number $n=246$ of candidate terms (\supp Section B).

For the flow equation, the usual form assumed for model-discovery is Navier-Stokes-like, with the time-derivative on the left side and rest of the terms on the right side \cite{Rudy2017,Reinbold2019,Reinbold2020,Reinbold2021}. However, this approach is not appropriate for our system. Since it is in the low Reynolds number regime \cite{Giomi2015,Doostmohammadi2017}, the significance of the time-derivative term itself needs to be investigated. 
Indeed, the active nematic flow has been modeled using pure Stokes \cite{Varghese2020,Zhou2021,Duclos2020}, unsteady Stokes \cite{Giomi2012, Giomi2015} as well as full Navier Stokes \cite{Giomi2013,Giomi2014,Doostmohammadi2018,Shendruk2017,Chandragiri2019,Chandragiri2020,Thampi2013,Thampi2014,Thampi2015,Thampi2016} formulations. While these approaches have been compared numerically \cite{Koch2021}, there has yet to be a definitive indication of the contributions of the inertial terms for this system. Since the viscous forcing is guaranteed to exist in this regime, we assume a form
\begin{equation}
    \nabla^2 \vb{u} = c_0 \partial_t \vb{u} + \sum_i c_i \vb{H}_i(\q, \vel,\nabla\q, \nabla\vel, \ldots) 
    \label{eq:uform}
\end{equation}
with $\nabla \cdot \vb{u} = 0$, and the time-derivative on the \emph{right hand side} so that its contribution can be evaluated. For instance, the lowest order symmetry-allowed `active stress' in the flow equation is the well-known $-\alpha \vb{Q}$, with $\alpha>0$ being the extensile `activity' \cite{Doostmohammadi2018,AditiSimha2002}. In our model form, this gives a general flow equation:
\begin{equation*}
\nabla^2\vb{u} = c_0 \partial_t \vb{u} + c_1 \vb{u}\cdot\nabla\vb{u} + c_2 \nabla P + c_3 \nabla\cdot\vb{Q} + \ldots    
\end{equation*}
which can be captured by our method. In this form, the coefficient $c_3$ corresponds to the ratio of the activity to the viscosity, $\alpha/\eta$. In practice, we use the stream function formulation of the flow equation to eliminate the pressure, which is not an observable in the experiments ~\cite{Rudy2017,Reinbold2019}. 

We perform model discovery from the data as follows.
Setting $N_x$, $N_y$, and $N_t$ as the number of measurements in the two spatial dimensions and time respectively, we randomly select $m$ of the total $N_x N_y N_t$  space-time points. At each selected space-time point we evaluate a linear system, \eg for the $Q_{xx}$ equation, $(\partial_t Q_{xx})_{m\times 1} = F_{m\times n} \cdot \vec{a}_{n \times 1}$.

The derivatives must be computed numerically, which amplifies noise in the data. 
To mitigate noise, we use two different approaches. In the \emph{integral formulation}, for each of the $m$ selected space-time points and $n$ terms, we compute a local average in space and time in a small window (e.g. \si{5x5x5} pixels) \cite{Alves2020}. This approach is effective for model discovery, but leads to inaccurate parameter estimates for the flow equation because the noise mitigation is not sufficient to counter noise amplification by the high-order derivatives in that equation.

To obtain more powerful noise mitigation at the cost of additional analytical effort, we adapt a \emph{weak formulation} of the PDE regression problem ~\cite{Reinbold2019,Reinbold2020,Reinbold2021}. Briefly, we fit the data to the weak form of Eq.~\eqref{eq:uform}:
\begin{equation}
    \int_{\Omega_k} \vb{w} \cdot \left[ \nabla^2 \vb{u} = c_0 \partial_t \vb{u} + c_1 \vb{u}\cdot\nabla\vb{u} +  \ldots \right]
    \label{eq:uformweak}
\end{equation}
By choosing an appropriate test function $\vb{w}$, we can move the derivatives from the noisy experimental data to the exact test functions, and also integrate out latent variables using integration by parts (see \supp Section B for details, the terms include in the library are available in Table~\ref{table:flow_library}).

Next, we seek optimal fits to these equations with the minimum number of non-zero terms, thus yielding an interpretable model that accurately describes the data but avoids overfitting. To this end, we perform optimization using Ridge regression (least-squares gives similar results), and then eliminate the least important terms one by one to obtain a hierarchy of models~\cite{Alves2020}. The optimal model is then determined by comparing the trade-off between the accuracy ($R^2$ of the fit) and the complexity (number of non-zero terms in the model) \cite{Alves2020}.

We first test our framework by recovering governing equations from data generated by simulating active nematic hydrodynamics. To generate the data, we consider two qualitatively different models for flow: one is purely Stokesian with substrate friction, and the other is unsteady Stokes flow, which omits the convective term but keeps $\partial_t \vb{u}$. 
Although both models capture the primary physics of active nematics, they test the algorithm with different types and numbers of terms \cite{Doostmohammadi2018,Duclos2020,Zhou2021,Varghese2020,Giomi2012,Giomi2015}. 
We solve these equations numerically and extract the steady state Q-tensor and velocity values (\supp Section C and Table~\ref{table:parameters}). To test noise mitigation, we add 5\% synthetic noise to the simulation data, and apply the integral formulation to the alignment equation and the weak formulation to the flow equation. 

\begin{figure}
    \centering
    \includegraphics{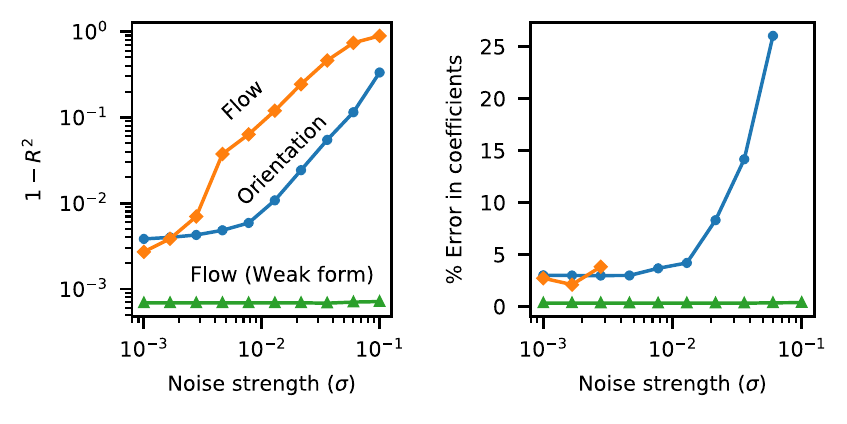}
    \caption{Performance on simulation data as a function of added noise. (left) $(1-R^2)$ as a function of noise strength for the orientation equation (blue circles), flow equation (orange diamonds) and the flow equation in the weak form (green triangles). The first two are obtained using the weak formulation. (right) The corresponding average \% error in the fitted coefficients. For the points absent on the plot, the obtained optimal model did not match with the ground truth. A $5\times5\times5$ volume was used for the integral formulation, whereas a $205\times 205\times 65$ window was used for the weak formulation.}
    \label{fig:noise_benchmark}
\end{figure}

While eliminating terms one by one, we obtain the $R^2$ value at each model hierarchy. We plot the \emph{optimality curve} as the logarithm of the Fraction of Variance Unexplained (FVU), or ($1-R^2$) as a function of the number of non-zero terms left in the model. We define the optimal number of terms $\nStar$ as the $n$-value at which the second derivative of the curve is highest, indicating the largest drop in $\log (\mathrm{FVU})$ (Fig.~\ref{fig:sindy_sim}). The equation used for the alignment tensor Eq.~\eqref{eq:alignment}, in particular for $Q_{xx}$, has 10 terms (\supp Section C), while the Stokes and Unsteady Stokes equations both have 2 terms. In all these cases, the framework returns the correct equation with very small errors in the identified coefficients (Fig.~\ref{fig:sindy_sim} and \supp  Fig.~\ref{fig:box_size_benchmark}). Thus, we estimate important phenomenological parameters directly from the data, including the the activity level $\alpha$, bending modulus $K$, flow alignment coupling $\lambda$, and bulk free energy coefficients $a_2$ and $a_4$. We benchmark the noise mitigation strategies by performing the same protocol with increasing window sizes (\supp Fig.~\ref{fig:noise_benchmark}) and varying levels of synthetic noise (\supp Fig.~\ref{fig:noise_benchmark}). The results show that the integral formulation with a small window size of a few pixels is suitable for the alignment equation, while the weak formulation with a large window size, comparable to the system size, is better for the flow equation.

\begin{figure}
    \centering
    \includegraphics{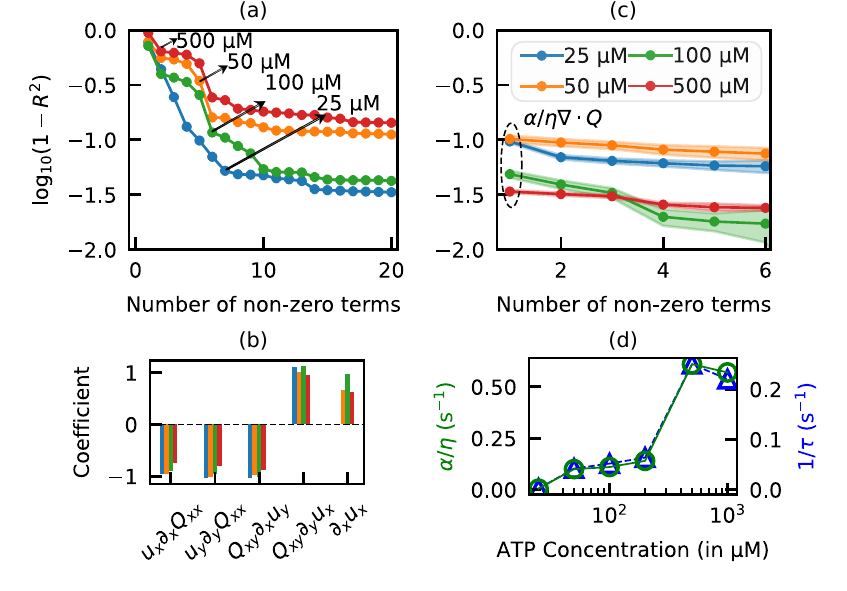}
    \caption{\boldf{Discovering active nematic hydrodynamics from experimental data.} (a) Optimality curves (($1-R^2$) vs number of non-zero terms) for the $Q_{xx}$ equation from experiments with various ATP concentrations. The shoulder at $n\sim 6$ contains flow coupling terms --- convection, rotation and flow alignment. Higher order flow alignment terms show up at higher $n$. The blue, orange, green and red colors correspond to \um{25}, \um{50}, \um{100} and \um{500} ATP respectively, in (a-c). (b) Values of the coefficients of key flow-coupling terms appearing in the optimal models for various ATP concentrations. (c) Optimality curves for the weak form flow equation. The activity term $\nabla\cdot\vb{Q}$ dominates, suggesting Stokesian dynamics. (d) The fitted coefficient of the activity term, $\alpha/\eta$, as a function of the ATP concentration (green circles). This quantity compares well with the inverse of the velocity correlation time (blue triangles), suggesting that $\alpha/\eta$ corresponds to a relevant timescale in the system.}
    \label{fig:sindy_expt}
\end{figure}

Next, we perform model discovery on orientation and velocity data extracted from fluorescence images of microtubule-based active nematic (\supp Section D). We varied the ATP concentration, which determines the motor stepping speed and thus determines the structure and dynamics of active nematics. Figs.~\ref{fig:sindy_expt}a,c show the optimality curves for the alignment and flow equations respectively at various ATP concentrations. In most cases, we obtain the optimal model:
\begin{align*}
\partial_t \vb{Q} \sim \ &-\vb{u}\cdot\nabla\vb{Q} - (\vb{\Omega}\cdot\vb{Q} - \vb{Q}\cdot\vb{\Omega}) \\ 
&+ \vb{E} - 2(\vb{Q}\colon\nabla\vb{u}) \vb{Q} \\ \tageq
\eta \nabla^2\vb{u} \sim \ &+ \alpha \nabla\cdot\vb{Q} + \nabla P 
\label{eq:optimalModel}
\end{align*}
As tests of consistency, the alignment equation recovers Galilean invariance from the data: the convective and co-rotational derivatives have coefficients of $\sim 1$ (Fig.~\ref{fig:sindy_expt}b). Furthermore, the flow alignment parameter, $\lambda \sim 1$ (Fig.~\ref{fig:sindy_expt}b), is consistent with the theoretical result for the high aspect ratio $a \gg 1$ of the microtubules, $\lambda = (a^2-1)/(a^2+1) \to 1$ \cite{Maitra2018} \footnote{A higher accuracy result for the Q-tensor equation was obtained using a dataset with a small field of view with a few defects per frame (see \supp Section E). That dataset however doesn't give a good $R^2$ for the flow due to the limitation on the window size.}. Interestingly, the terms arising from the bulk liquid crystal free energy (see Eq.~\eqref{eq:Qtheory}) are absent, a finding that supports a previous model~\cite{Thampi2015}. Elastic free energy terms such as $ \nabla^2 \mathbf{Q}$ are also absent. These results indicate that the alignment dynamics of microtubule-based active nematics are dominated by flow coupling. In comparison, contributions from the free energy dissipation to the dynamics are negligible. The optimality curves for the flow equation are almost flat (Fig.~\ref{fig:sindy_expt}c), showing that the activity term $\alpha/\eta \nabla\cdot\vb{Q}$ alone dominates the  dynamics. The inertial terms  are absent (not appearing until $\nStar\sim5$), indicating that the Stokes flow approximation accurately describes the experimental active nematic. Finally, the absence of the substrate friction term indicates that the flows are largely unscreened.

\begin{figure}
    \centering
    \includegraphics{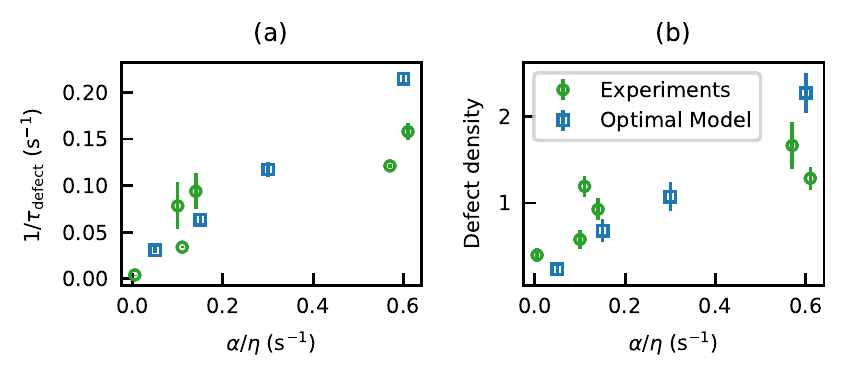}
    \caption{Comparison of the simulated optimal model with the experimental data. (a) Inverse of the lifetime of $+1/2$ defects plotted as a function of the value of $\alpha/\eta$ obtained from the optimal model (green circles) and used in the simulations (blue squares). The height of the errorbars is twice the standard error of mean. (b) Defect density plotted against the value of $\alpha/\eta$ obtained from the optimal model (green circles) and used in the simulations (blue squares). The density in the simulations is scaled by a constant. The height of the errorbars is twice the standard deviation.}
    \label{fig:comparison}
\end{figure}
The terms in the optimal model for $\q$ have dimensionless coefficients. Thus, we cannot extract length scales or time scales from the Q-tensor equation, or provide quantitative estimates for dimensional experimental parameters characterizing $\q$ dynamics \footnote{We can force the framework to estimate dimensional parameters by constraining the regression procedure to include specific terms, while performing sparse regression on the remaining terms. For example, by forcing a term $K \nabla^2 Q_{xx}$ (see Eq.~\eqref{eq:Qtheory}), we obtain a value for the elastic modulus  of $K \sim $ \SI{1}{\micro\meter\squared\per\second}. However, because this term has a negligible contribution to the dynamics of $\q$, the quantitative accuracy of this estimate may be limited.}, which in the context of the discovered optimal model, is an emergent consequence of the dynamics. 
The fit to the flow equation provides a direct estimate of the scaled activity parameter $\alpha/\eta$, an intrinsic time scale \cite{Giomi2011}, as a function of the ATP concentration (Fig.~\ref{fig:sindy_expt}d) \footnote{Independent measurements of the viscosity such as reported in \cite{Guillamat2016} can be then used to estimate $\alpha$, the strength of the active force}. 
Determining the relationship between activity and experimental control parameters has been a significant challenge~\cite{Lemma2019}. To test this estimate against another observable, we compare this time scale to the velocity autocorrelation time $\tau$, defined as $\bar{C}_{v}(\tau)=1/e$, with the autocorrelation function 
$\bar{C}_{v}(t) = \expval{ \expval{ \vb{u}(\vb{r},t'+t) \cdot  \vb{u}(\vb{r},t') }_{t'} / \expval{ \vb{u}(\vb{r},t') \cdot  \vb{u}(\vb{r},t') }_{t'} }_{\vb{r}}$. These observables closely agree over the whole range of  ATP concentrations (Fig.~\ref{fig:sindy_expt}d). 

Finally, we test the optimal model by performing simulations of Eq.~\eqref{eq:optimalModel} \footnote{For numerical stability, we include the $K \nabla^2 \vb{Q}$ term in the $\vb{Q}$ equation with $K=1$} and comparing the mean defect lifetime and defect density as a function of $\alpha/\eta$ (Fig.~\ref{fig:comparison}). Remarkably, the defect lifetimes for experiments and simulations align well without any pre-factors (Fig.\ref{fig:comparison}a), indicating a scale-free mapping between the time units. The defect densities from experiment and simulation also match, within a constant scaling factor that cannot be directly estimated since we lack a length scale (Fig.\ref{fig:comparison}b) \footnote{We re-scale the defect density so that the simulation value for $\alpha/\eta=0.3$ falls on the linear fit of the experimental values. }.

In summary, we have applied a data-driven PDE discovery method 
to identify equations governing both the orientational dynamics and the activity-driven flows
of microtubule-based active liquid crystals.
The optimal model is surprisingly minimal. It demonstrates that: (1) flow coupling dominates the orientational dynamics, and (2) the lowest-order active stress, proportional to the local orientational order, together with the vanishing Reynolds number limit describe the flow. The identified equations enable mapping between key model parameters and experimental control variables, including the elusive relationship between the magnitude of activity and ATP concentration.
Through comparison of several noise reduction approaches on the experimental data, as well as extensive benchmarking against simulated data, we have identified an approach to model discovery which is highly robust against experimental noise. We have delineated associated requirements for experimental data acquisition; i.e., requirements on the amount of data as well as its temporal and spatial resolution. The approach is general, and can be applied to a wide variety of active matter systems as well as other experimental and computational observations of dynamics.

\begin{acknowledgements}
We acknowledge support from the National Science Foundation (NSF) DMR-1855914, OAC-2003820 and the Brandeis Center for Bioinspired Soft Materials, an NSF MRSEC (DMR-2011486), as well as computing resources through NSF XSEDE allocation TG-MCB090163 (Stampede and Comet) and the Brandeis HPCC which is partially supported by the NSF through DMR-MRSEC 2011486 and OAC-1920147. We thank Link Morgan for providing early experimental data for testing, Saaransh Singhal for providing the simulation data for the unsteady Stokes equation, Peter J. Foster for providing feedback on the manuscript, and Michael M. Norton and Seth Fraden for valuable discussions.
\end{acknowledgements}
\section*{\label{sec:theory}Appendix A: Continuum Theory of Active Nematics}

The theory of active nematic suspensions builds on equilibrium nematic hydrodynamics \cite{degennes_prost_1993,beris_edwards_1994} and extends it to include non-equilibrium `active stresses' in the fluid \cite{Ramaswamy2010,Marchetti2013,Giomi2012,Giomi2014b,Thampi2014,Gao2015,Giomi2015,Maitra2018,Norton2018,Hemingway2016,Ngo2014}.
The most generic models of active nematics have considered the dynamics of not only the orientation and the velocity, but also of the concentration \cite{Giomi2012,Giomi2013,Giomi2014} and/or density \cite{Thampi2015}. In this work, we assume uniform density and concentration everywhere (a reasonable assumption for dense 2D bulk systems, while we do believe that allowing the density/concentration to vary will only improve our results). 
The orientational order is often defined using a tensor order parameter $\vb{Q} = S [\vb{n}\otimes\vb{n}-(1/2)\vb{I}]$. This definition has the $\vb{n}\to-\vb{n}$ nematic symmetry built-in, and also provides a scalar order parameter $S(\vb{r},t)$ that represents the magnitude of the orientational order at the given location.
For equilibrium systems, the free energy near the isotropic-nematic transition takes the form
\begin{equation}
    \mathcal{F} = \frac{a_2}{2} Q^2 + \frac{a_4}{4} Q^4 + \frac{K}{2}(\nabla Q)^2
\end{equation}
Here, $a_2 = (1-\rho)$ and $a_4 = (\rho+1)/\rho^2$. The density $\rho$ controls the transition from the isotropic ($\rho < 1$) to the nematic ($\rho > 1$) state. If  $\rho = \rho_0 > 1$, the minimum of the free energy is in an ordered state, with $S_{eqm}= \sqrt{-2 a_2/a_4}$. $K$ is the elastic modulus of the nematic, assuming equal moduli for splay and bend deformations. 
For the alignment tensor, the dynamical equations of motion are:
\begin{equation}
    \label{eq:q_tensor}
    \frac{\mathrm{D}\vb{Q}}{\mathrm{D}t} = \partial_t \vb{Q} + \vb{u}\cdot\nabla\vb{Q} + (\vb{\Omega}\cdot\vb{Q} - \vb{Q}\cdot\vb{\Omega}) = \lambda \vb{E} + D_r \vb{H}
\end{equation}
The left hand side corresponds to the co-moving co-rotational derivative of the Q-tensor, with $\vb{\Omega} = 1/2(\nabla \vb{u} - (\nabla\vb{u})^T)$ and $\vb{E} = 1/2(\nabla \vb{u} + (\nabla\vb{u})^T)$. $\lambda$ is the flow alignment parameter, with $\lambda \vb{E}$ being the lowest order contribution to the flow alignment. The next leading order term $-2\lambda (\vb{Q}\colon\vb{E}) \vb{Q}$ \footnote{The second order term --- $\lambda\{ \vb{Q}\cdot\vb{E}+\vb{E}\cdot\vb{Q} - 2/d \Tr(\vb{Q}\cdot\vb{E})\vb{I} \}$ \cite{Doostmohammadi2018,Varghese2020}---vanishes identically in 2D} is also often included \cite{Doostmohammadi2018}. Lastly, $\vb{H} = -\delta \mathcal{F}/\delta \vb{Q}$ and $D_r$ is the coefficient of rotational diffusion. This term is the free energy contribution, while all the others come from the coupling to the underlying flow.  

The corresponding fluid flow can be described by the incompressible Navier-Stokes (NS) equations. These are augmented with the (passive) back-flow $\sigma_p$ due to the coupling to the nematic. In addition, the lowest order \text{non-equilibrium} nematic stress takes the form $-\alpha \vb{Q}$, where $\alpha$ indicates the strength of the active forces, or ``activity". Adding this contribution in, we get the \emph{active} nematic fluid equation:
\begin{equation}
    \rho(\partial_t \vel + \vel\cdot\nabla\vel) = \eta \nabla^2 \vel - \nabla P + \nabla\cdot(\sigma_p - \alpha \vb{Q})
    \label{eq:ns}
\end{equation}
with $\rho$ being the density and $\eta$ being the viscosity. 

For thin 2D samples such as the ones considered in this \paper, Eq.~\eqref{eq:ns} needs to be integrated along the height $h$ of the confinement to get a quasi-2D equation. Integrating the viscous term gives an effective linear friction $-\Gamma \vb{u}$, with $\Gamma \sim \eta/h^2$ \cite{Maitra2018}. 

Since the flows have a low-Reynolds number, it is common to neglect the non-linear convection term \cite{Giomi2012,Giomi2014}, and often the entire inertia term \cite{Duclos2020,Zhou2021,Varghese2020}. This is the Stokes limit. The theoretical consequences of varying the advective inertia and substrate friction have been the subject of numerical studies \cite{Thampi2014,Koch2021}.

Additionally, a higher order active stress $\nabla (\vb{Q}\cdot\nabla\cdot\vb{Q})$ in 3D can give rise to a non-equilibrium active force $\vb{Q}\cdot\nabla\cdot\vb{Q}$ in 2D, which can be expected to have a similar magnitude as the primary active force $-\alpha\nabla\cdot\vb{Q}$ \cite{Maitra2018}.  

This brief sketch of active nematics models shows the wealth of information that a data-driven method like SINDy could uncover when applied to experimental systems. For the numerical model in this \paper, we use the Stokes limit described above. Further, we set $\sigma_p=0$ as it has been observed numerically that active stresses dominate the passive stresses \cite{Thampi2016,Doostmohammadi2018}. 
We include the substrate friction as it adds another degree of freedom that our framework has to identify. With this, we get our dynamical equation for the fluid:
\begin{align}
    \label{eq:velocity}
    \eta \nabla^2 \vb{u} - \Gamma \vb{u} - \nabla P - \nabla\cdot (\alpha\vb{Q}) &= 0 \\
    \nabla \cdot \vb{u} &=0 \nonumber
\end{align}
The Q-tensor equation is unaltered by activity because the active stresses come into picture through the flow coupling. 
\section*{Appendix B: Library creation and benchmarking}
Each data-set is pre-processed such that the fields $Q_{xx}$, $Q_{xy}$, $u_x$ and $u_y$ lie on the same $N_x \times N_y$ grid. With $N_t$ such measurements in time, we get $N_x \times N_y \times N_t$ values for each of the field.

\paragraph{Integral Formulation} We begin by creating a database of terms containing the fields and their derivatives. We compute the time and space derivatives numerically using a central difference scheme. (Since $\nabla\cdot \vb{u} = \partial_x u_x + \partial_y u_y = 0$, we discard $\partial_y u_y$ from our database.)
To form the library, we then make all multiplicative combinations of these terms with a total functional order up to $f$ and a total gradient order up to $d$. 
We put further limits on the function and gradient order of $\vb{u}$ appearing in the terms.
Thus, we specify two pairs, ($f_t$ , $d_t$) for the overall constraint,  and ($f_u$, $d_u$) for further constraint on the terms involving velocity. This allows us to make the computation more tractable. Motivated by theoretical models discussed in the \supp Section A, we use ($f_t =3, d_t =2$) and ($f_u =1, d_u =1$) for the $\vb{Q}$ tensor equation (Eq.~\eqref{eq:Qform}) and ($f_t =2, d_t =2$) for the flow equation (Eq.~\eqref{eq:uform}). In this approach, we don't make any simplifying assumptions about the terms : all terms appear in an ``unfolded" form, with the derivatives and inner products expanded out.

Since our data has two spatial and one time dimensions, we have a large number of data-points. Hence, we compute the library terms only on a sub-sample of the data \cite{Rudy2017} : we randomly select $m=5000$ points from the $N_x \times N_y \times N_t$ grid and compute the local average of the terms near the points using a small (\eg \si{5x5x5} pixels) averaging window \cite{Alves2020}. 

\paragraph{Weak Formulation} For the fluid flow, we assume a generalized flow equation
\begin{equation*}
    \int_{\Omega_k} \vb{w} \cdot \left[ \nabla^2 \vb{u} = c_0 \partial_t \vb{u} + c_1 \vb{u}\cdot\nabla\vb{u} +  \ldots \right]
\end{equation*}
\begin{table}
\centering
\begin{tabular}{|c||c|c|} 
 \hline
 Number & Term \\ 
 \hline\hline
 1 & $\partial_t \vb{u}$  \\ 
 \hline
 2 & $\vb{u}\cdot\nabla \vb{u}$ \\
 \hline
 3 & $\nabla\cdot\vb{Q}$ \\
 \hline
 4 & $\vb{Q}\cdot\nabla\cdot\vb{Q}$  \\ 
 \hline
 5 & $\vb{u}$  \\ 
 \hline
 6 & $\vb{Q}\cdot\vb{u}$  \\ 
 \hline
 7 & $\Tr (\vb{Q}^2) \vb{u}$\\ 
 \hline
\end{tabular} 
 \caption{Terms appearing in the library for the flow equation.\label{table:flow_library}}
\end{table}
Using a similar ``unfolded" form for the library would generate a lot of terms, and it is not feasible to perform the integration by parts for all of them. Hence, we create the library by hand with a judicious choice of terms, listed in Table~\ref{table:flow_library}. Here,  $\vb{w}$ is a vector test function and the integration domain $\Omega_k$ is a rectangular box of size $2H_x \times 2H_y \times 2H_t$, centered at $(x_k,y_k,t_k)$ \cite{Reinbold2020}. Following \cite{Reinbold2020}, we use the test function 
\begin{equation}
    \vb{w} = \nabla \times (\psi \hat{z})
    \label{eq:w_def}
\end{equation}
with 
\begin{equation}
    \psi = \sin(\pi \ubar{t})(\ubar{x}^2-1)^p (\ubar{y}^2-1)^p
    \label{eq:psi_def}
\end{equation}
where the underbar represents the rescaled variables  $\ubar{x}=(x-x_k)/H_x$, $\ubar{x}=(y-y_k)/H_y$ and $\ubar{x}=(t-t_k)/H_t$. Eq.~\eqref{eq:w_def} implies that $\nabla\cdot\vb{w}=0$, which facilitates the elimination of pressure (see below). The functional form of Eq.~\eqref{eq:psi_def} guarantees that $\vb{w}$ and its derivatives vanish at the domain boundaries given a sufficiently large value of $p$. This is useful for the integration by parts, which goes as follows:

\begin{alignat}{3}
    u^k_0 &= \int_{\Omega_k} \! \vb{w} \cdot \nabla^2 \vb{u} \ \dd \Omega &&= \int_{\Omega_k} \! \vb{u} \cdot \nabla^2 \vb{w} \ \dd \Omega\\
    u^k_1 &= \int_{\Omega_k} \! \vb{w} \cdot \partial_t \vb{u} \ \dd \Omega &&= -\int_{\Omega_k} \! \vb{u} \cdot \partial_t \vb{w} \ \dd \Omega\\
    u^k_2 &= \int_{\Omega_k} \! \vb{w} \cdot (\vb{u} 
    \cdot \nabla \vb{u}) \ \dd \Omega &&= -\int_{\Omega_k} \! \vb{u}\cdot (\vb{u}\cdot\nabla) \vb{w}  \ \dd \Omega 
\end{alignat}
The activity term integrates as follows,
\begin{equation}
    u^k_3 = \int_{\Omega_k} \! \vb{w} \cdot (\nabla\cdot\vb{Q}) \ \dd \Omega= -\int_{\Omega_k} \! \vb{Q} \colon (\nabla\vb{w}) \ \dd \Omega
\end{equation}
while the pressure term integrates out to zero as follows:
\begin{equation}
\int_{\Omega_k} \! \vb{w}\cdot\nabla P \ \dd \Omega = -\int_{\Omega_k} \! P \nabla\cdot\vb{w} \ \dd \Omega = 0     
\end{equation}
The derivative on $\vb{Q}$ in term 4 in Table~\ref{table:flow_library} cannot be fully transferred to $\vb{w}$ because of the non-linearity, so we have to integrate that term directly. Similarly, terms \numrange{5}{7} are integrated directly as they do not contain any derivatives.

We benchmark these methods against varying noise levels as well as varying window sizes. To find the appropriate window size, we benchmark with a numerical data of size $512\times 512 \times 500$ with $5\%$ added noise. We measure the $R^2$ value of the fit as well as the average \% error in the coefficients in the optimal model (if found correctly) as a function of the integration window size (Fig.~\ref{fig:box_size_benchmark}). For the integral formulation (Fig.~\ref{fig:box_size_benchmark} left), we find that a small window size is sufficient to mitigate the noise. For the experimental dataset results in Fig.~\ref{fig:sindy_expt}, we use a window size of $25\times 25\times 25$ pixels, or \SI{73}{\micro\meter}$\times$\SI{73}{\micro\meter}$\times$\SI{12}{\second}. In the weak formulation, larger window sizes are needed to better sample the test function \cite{Reinbold2020,Reinbold2021}. We find that a window that is almost as large as the field of view in space, and $\sim 5$ times the velocity correlation time in the time dimension \cite{Reinbold2021} works well. We perform a similar analysis with the experimental data-sets to choose the window size. To avoid self-selection, we use the largest of the correlation times from the data-sets to set the window size. In Fig~\ref{fig:sindy_expt}, we use a window size of $205 \times 205 \times 605$ pixels, or $\sim$ \SI{600}{\micro\meter}$\times$\SI{600}{\micro\meter}$\times$\SI{300}{\second}. As this window size is large, we take $m=50$ measurements for the weak form. 

We now investigate the same two parameters, namely the $R^2$ and the error in coefficients, but with varying noise levels (Fig.~\ref{fig:noise_benchmark}). Our analyses, and indeed our results, indicate that the integral formulation is sufficient for the Q-tensor equation, while the weak formulation works adequately for the flow equation. 

\begin{algorithm}
\label{algo:hridge}

\KwResult{$\vb{A}, r2s = \texttt{HRidge}(\vb{F},\vb{\dot{X}},\lambda_2$)}

$m$ = size($\vb{F}$)[2] \tcp*{Total number of terms}
$F_n = \texttt{norm}(\vb{F})$ \tcp*{Column-wise norm}
$\vb{F}_0 = \vb{F}/F_n$ \tcp*{Normalize the terms for comparison}
$\vb{A} \leftarrow$ Vector(size: $m\times m$) \tcp*{Store optimal model at all hierarchies}
$r2s \leftarrow$ Vector(size: $m$) \tcp*{Store model accuracies at all hierarchies}
$\vb{a} = \argmin_{\vec{a}} \left( \lnorm{\dot{\vb{X}}-\vb{F}_0 \cdot \vb{a}}_2 + \lambda_2 \lnorm{\vb{a}}_2\right)$  \tcp*{Initial guess using Ridge}
$\vb{A}[:, m] = \vb{a} / F_n$ \tcp*{Store un-normalized coefficients}
\BlankLine
$r2s[m] =\texttt{rsquared}(\vb{F}_0,\vb{\dot{X}},\vb{a})$\\
\While {$len(\vb{a})>1$}{
$j_0 = \argmin_{j} |a_j| $ \tcp*{Find the position of the smallest coefficient} 
$\vb{a} = \vb{a} \setminus \{ a_{j_0}\}$  \tcp*{Remove smallest coefficient} 
$\vb{F}_0 = \vb{F}_0 \setminus \vb{F}_0[:,j_0]$ \tcp*{Remove vector corresponding to it} 
$\vb{a} = \argmin_{\vec{a}} \left( \lnorm{\dot{\vb{X}}-\vb{F}_0 \cdot \vb{a}}_2 + \lambda_2 \lnorm{\vb{a}}_2\right)$\;
$k = len(\vb{a})$\;
$\vb{A}[:, k] = \vb{a} / F_n $\;
$r2s[k] =\texttt{rsquared}(\vb{F}_0,\vb{\dot{X}},\vb{a})$\;
}


\caption{Hierarchical Ridge Regression (HRidge)}
\end{algorithm}

\section*{Appendix C: Continuum Simulations}
\label{si:simulations}

For the Q-tensor, we use the simple form (in non-dimensionalized units)
\begin{equation}
    \partial_t \vb{Q} + \nabla\cdot(\vb{u}\vb{Q}) + (\vb{\Omega}\cdot\vb{Q} - \vb{Q}\cdot\vb{\Omega})  =
    \lambda \vb{E} + \vb{H} \label{eq:alignment}
\end{equation}
with 
\begin{align*}
    \Omega_{ij} &= \frac{1}{2}(\partial_i u_j - \partial_j u_i) \\
    E_{ij} &= \frac{1}{2}(\partial_i u_j + \partial_j u_i) \\
    H_{ij} &= (-a_2-a_4 Q_{kl}Q_{lk}) Q_{ij} + K \partial_k \partial_k Q_{ij}
\end{align*}
where $\lambda$ is the flow alignment parameter, $K$ is the bending modulus and $a_2<0, a_4>0$ drive the system to the nematic phase.
The two different flow equations used are
\begin{align}
    \eta\nabla^2 \vb{u} &= \nabla P + \Gamma \vb{u} + \alpha \nabla\cdot \vb{Q} \quad \text{(Stokes)} \label{eq:stokes}\\
    \rho\partial_t \vb{u} &= \eta \nabla^2 \vb{u} - \nabla P - \alpha \nabla\cdot \vb{Q}  \quad \text{(Unsteady Stokes)}\label{eq:unsteadystokes}\\
    \nabla\cdot\vb{u} &= 0 \label{eq:incomp}
\end{align}
Here, $\eta$ is the viscosity, $\alpha$ is the activity and $\Gamma$ is the substrate friction. We set $\rho=1$ for the unsteady Stokes equation.

For the Stokes equation Eq.~\eqref{eq:stokes}, we use a semi-implicit finite difference time stepping scheme based on a convex splitting of the nematic free energy \cite{Zhao2016,Varghese2020}. To solve the Stokes equation with incompressibility, we implement a Vanka type box smoothing algorithm on a staggered grid \cite{Vanka1986,Varghese2020}. The solution at each time-step is found using  Gauss-Seidel relaxation iterations, and the rate of  convergence to the solution is accelerated by using a multigrid method \cite{Varghese2020}. The simulation codes are all in-house and are written in C. We solve the equations in a square domain of size $200\times200$ (in simulation units) with periodic boundary conditions. We sample the evolution at a time step of $1$ (in simulation units) on a rectangular grid with $\dd x \sim 0.4$. For the unsteady Stokes, the coupled equations are numerically solved in a forward time centred space (FTCS) scheme with an explicit time-stepped predictor-corrector method.
The incompressible unsteady Stokes equations use a modified Chorin's projection method \cite{Hermle2017} to compute the velocities. The data was sampled at a time-step of $0.02$ on a rectangular grid with $\dd x \sim 0.2$. The parameters used in both the cases are documented in Table~\ref{table:parameters}.

The equation for $Q_{xx}$ has 10 terms:
\begin{align*}
    \partial_t Q_{xx} = &-u_x\partial_x Q_{xx}  
    - u_y\partial_y Q_{xx}  
    -Q_{xy}\partial_x u_y  
    +Q_{xy}\partial_y u_x \\
    &+ \lambda \partial_x u_x 
    - a_2 Q_{xx} 
    -2 a_4 Q_{xx}^3
    -2 a_4 Q_{xy}^2 Q_{xx} \\
    &+ K \partial_x^2 Q_{xx} + K \partial_y^2 Q_{xx}
\end{align*}
The flow equations in the integral formulation are not used in the main text, but they would be in the form of vorticity equations (taking the curl of Eq.~\eqref{eq:stokes} and Eq.~\eqref{eq:unsteadystokes}):
\begin{align*}
    \eta\nabla^2 \omega &= \Gamma \omega + \alpha \partial_x^2 Q_{xy} - 2\alpha \partial_x\partial_y Q_{xx} - \alpha \partial_y^2 Q_{xy} \\
    \eta \nabla^2 \omega &= \partial_t \omega + \alpha \partial_x^2 Q_{xy} - 2\alpha \partial_x\partial_y Q_{xx} - \alpha \partial_y^2 Q_{xy}
\end{align*}
However, for the weak formulation, we get two terms each, after the pressure is eliminated as described in \supp Section B:
\begin{align*}
    \eta \int_{\Omega_k} \vb{w} \cdot \nabla^2\vb{u} &= \alpha \int_{\Omega_k} \vb{w} \cdot \nabla\cdot\vb{Q} + \Gamma \int_{\Omega_k} \vb{w} \cdot \vb{u} \\
    \eta \int_{\Omega_k} \vb{w} \cdot \nabla^2\vb{u} &= \alpha \int_{\Omega_k} \vb{w} \cdot \nabla\cdot\vb{Q} + \int_{\Omega_k} \vb{w} \cdot \partial_t \vb{u}
\end{align*}

\begin{table}
\centering
\begin{tabular}{|c||c|c|} 
 \hline
 Parameter & Stokes & Unsteady Stokes \\ 
 \hline\hline
 $\eta$ & 1 & 1  \\ 
 \hline
 $K$ & 1 & 1 \\
 \hline
 $\alpha$ & 0.3 & 4.0 \\
 \hline
 $a_2$ & -0.3 & -16  \\
 \hline
 $a_4$ & 1.36 & 32  \\ 
 \hline
\end{tabular} 
 \caption{Parameters used for the Stokes and Unsteady Stokes simulations.\label{table:parameters}}
\end{table}

\section*{Appendix D: Experimental Methods}

The active nematic samples were assembled following previously established methods \cite{Sanchez2012,DeCamp2015}. The active mix consisted of microtubules, kinesin motor clusters, depleting agent, and an ATP regeneration system. Tubulin was purified from bovine brain, labeled with Alexa 647 dye, and polymerized in the presence of GMPCPP \cite{Castoldi2003,Hyman1991}.  A truncated and biotinylated version of Kinesin-1 (K401-BCCP-HIS) was expressed in \textit{E. Coli} and purified using immobilized metal affinity chromatography~\cite{Subramanian2007}. Motor clusters were formed by incubating \SI{11}{\micro\liter} of the biotinylated kinesin (0.7 mg/ml) with \SI{5}{\micro\liter} of streptavidin (0.35 mg/ml) on ice for 30 min. Polyethylene glycol (35000 kDa, 1\%) was used to induce microtubule bundling. A biochemical regeneration system consisting of adenosine triphosphate (ATP, 25$\mu$M-1.4 mM), phosphoenol pyruvate (PEP, 26 mM), and pyruvate kinase/lactic dehydrogenase (PK/LDH) kept ATP concentration constant. Lastly, an oxygen scavenging system consisting of glucose (0.67 mg/ml), glucose oxidase (0.08 mg/ml), catalase (0.4 mg/ml), DTT (5.6 mM), and Trolox (2 mM) was used to minimize sample bleaching. The components of the active mix were combined in M2B buffer (80 mM PIPES, 1mM EGTA, 2mM MgCl2, pH 6.8). 

Flow chambers were created with Parafilm sandwiched between a glass slide and a coverslip. The glass slide was made hydrophobic with a Rain-X coating, and the coverslip was passivated with acrylamide coating \cite{Lau2009}. To assemble an active nematic, the chamber was first filled with HFE oil containing fluoro-surfactant (0.5\% w/w, RAN Biotech), followed by the active mix. The sample was sealed with UV glue (Norland optical adhesive). The active nematic sedimented to the oil-water interface and reached a steady state after about an hour, and was then imaged on a spinning disk confocal microscope using an Hamamatsu Orca-Fusion BT CMOS camera and 20$\times$ magnification. For each sample, a sequence of 10000 images was acquired at 2 frames/sec, except for the 25 $\mu$M ATP samples, for which 1000 frames were acquired at 0.1 frames/sec. 

\section*{Appendix E: Additional methods}

\subsection*{Orientation and velocity fields}
The orientation and velocity fields were computed simultaneously on the fluorescent images obtained from spinning disk confocal miscrocopy. The orientation fields were measured using an in-house structure-tensor-based code \cite{Rezakhaniha2012,Duclos2020} (written in MATLAB) on the fluorescence images. The molecular tensor ($\vb{n}\vb{n}-1/2\vb{I}$) was computed from the orientation data, and was then coarse-grained  with a Gaussian smoothing filter with $\sigma=10$ pixels to obtain the Q-tensor. The velocity fields were measured using particle-image velocimetry implemented by the MATLAB-based PIVLab software. The fields computed by PIVLab were post-processed using a Direct Cosine Transform - Penalized Least Squares (DCT-PLS) approach that validates the raw data, replaces the spurious and missing vectors and does some smoothing. \cite{Garcia2011}. The orientation fields are computed on a high resolution grid of $1152 \times 1152$ pixels, while the velocity fields are on a coarser grid of $71 \times 71$ pixels.  Therefore, both fields are interpolated on an intermediate grid of $256 \times 256$ pixels.

\subsection*{Defect detection and tracking}

To locate the defects, we compute a map of the signed winding number $w = 1/(2\pi) \oint \nabla\theta \cdot \dd{\vec{s}}$ at every point in space \cite{Kamien2002,Norton2018} with an integration ring of radius of 5 pixels. The winding number is zero everywhere except at the defect locations \cite{DeCamp2015,Ellis2017}. To eliminate spurious defects, we filter out regions with a non-zero winding number that are smaller than 60 squared pixels in area. 

Once the locations of the defects are obtained, the $+1/2$ defects are tracked using the open source software Trackpy \cite{Allan2021} using a \texttt{search\_range} value of 20 pixels. The trajectories thus obtained are further filtered with a threshold of minimum three frames of survival. 

\begin{figure*}
    \centering
    \includegraphics{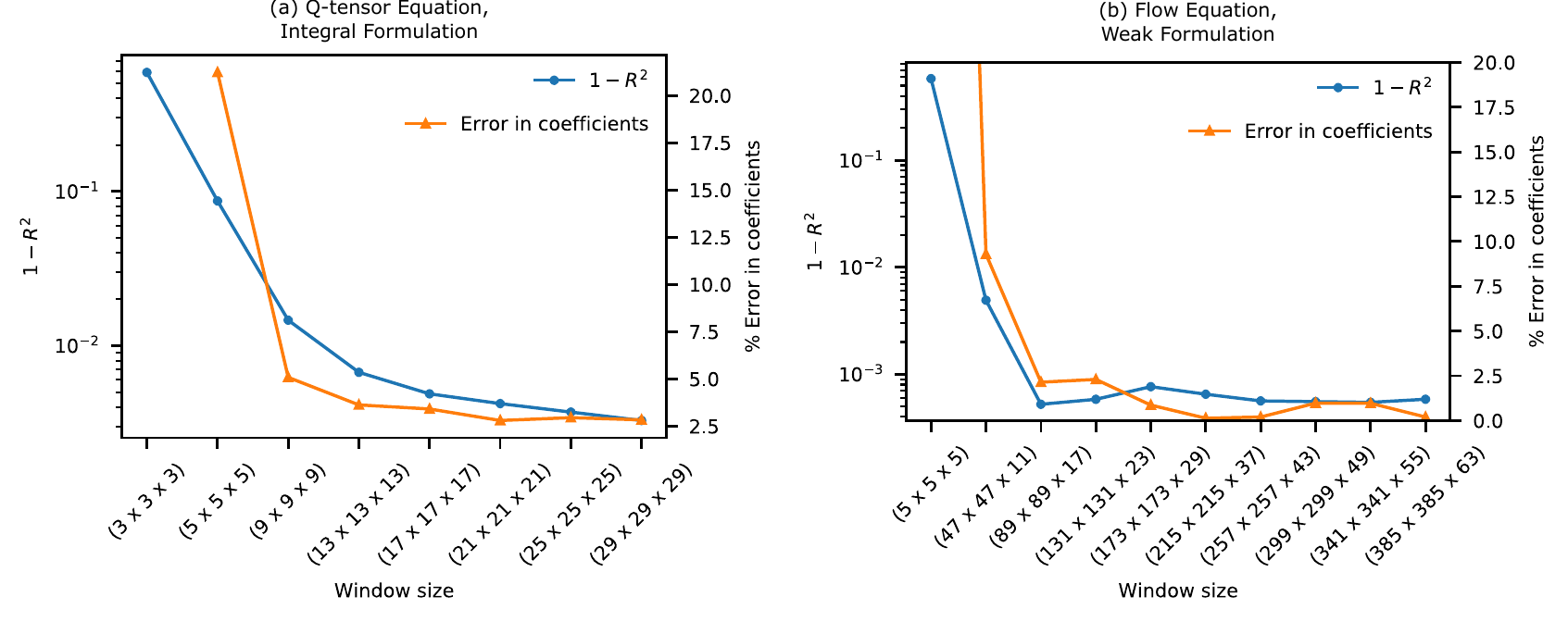}
    \caption{Performance on simulation data as a function of window size. $(1-R^2)$ (blue circles) and average \% error in coefficients (orange triangles) for the (left) Q-tensor equation using the integral formulation and (right) the flow equation using the weak formulation. The window sizes are listed in pixels, while the simulation data used had a domain size of $512\times 512 \times 500$ pixels.}
    \label{fig:box_size_benchmark}
\end{figure*}

\subsection*{Small field of view PolScope data}

An additional dataset was taken using a combination of LC-PolScope microscopy and dilute-labeled fluorescent MTs. With LC-PolScope microscopy, the orientation field is obtained from birefringence information of polarized light passing through the MT filaments that make up the nematic layer \cite{Oldenbourg2005} \cite{DeCamp2015}. This allows the orientation field to be measured on MTs that are not fluorescent. However, a small fraction of MTs in the sample were fluorescently labeled, and wide-field epifluorescence images were acquired simultaneously with the birefringence data. We believe that with wide-field microscopy, PIV on dilute-labeled MTs is more accurate than on fully-labeled MTs, due to the difficulty of detecting velocity in the direction of the elongated MT bundles. 

This sample was prepared as detailed in Appendix C at 1.4 mM ATP with two small differences: the chamber was created with double-sided tape and the hydrophobic slide was made with Aquapel. Additionally, the proteins used in this sample were from different preparations than the rest of the data in the paper. The LC-PolScope sample was imaged on a Nikon Ti Eclipse with Andor Neo camera, 20$\times$ 0.75 NA objective and 2 s frame interval. 

Using the integral formulation with a window size of $5\times 5\times 5$ pixels, we obtain the equation
\begin{align*} 
\partial_t Q_{xx} = &-(1.07 \ u_x \partial_x + 1.08 \ u_y \partial_y) \ Q_{xx} \\ &- (1.03 \partial_x u_y - 1.04 \partial_y u_x) \ Q_{xy}  \\ &+ 0.99 \partial_x u_x  \\
&- 2 Q_{xx} \{ 2.01 Q_{xx} \partial_x u_x  \\
&\quad + Q_{xy} (1.00 \partial_x u_y + 1.05 \partial_y u_x) \}
\end{align*} 
with a high $R^2$ value of 0.97. This provides further strong evidence for the discovered model. The flow analysis also yields a model similar to \eqref{eq:optimalModel}, but with a low $R^2$, due to the limitation on the window size due to the small field of view (see Fig.~\ref{fig:box_size_benchmark}).
\begin{figure}
    \centering
    \includegraphics[width=\linewidth]{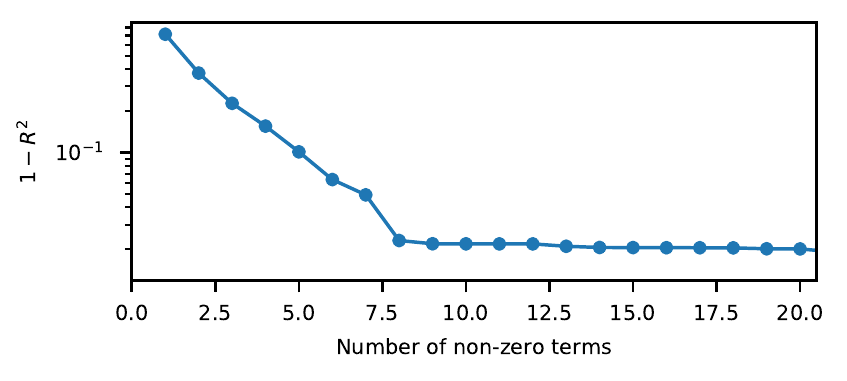}
    \caption{\boldf{Identifying Q-tensor Equation using PolScope measurements.} Optimality curve for the $Q_{xx}$ equation for a small field-of-view dataset in which the orientation is measured using PolScope microscopy. The $R^2$ at the shoulder is $\sim 0.97$.}
    \label{fig:q-tensor-polscope}
    
\end{figure}

\bibliography{main}

\end{document}